\documentclass[aip,reprint]{revtex4-1}

\usepackage{lipsum}
\usepackage{amsmath}
\usepackage{amssymb}
\usepackage{graphicx}
\usepackage{gensymb}
\usepackage{tabularx}
\usepackage{float}
\usepackage{xcolor}
\usepackage{mathtools}
\usepackage[capitalise]{cleveref}
\DeclareUnicodeCharacter{2009}{ }
\DeclareUnicodeCharacter{2192}{$\rightarrow$}
\DeclareUnicodeCharacter{2032}{'}
\DeclareUnicodeCharacter{03B1}{$\alpha$}
\DeclareUnicodeCharacter{03B3}{$\gamma$}
\restylefloat{table}
\graphicspath{ {Images/} }

\begin{document}
	\title{The Effect of Areal Density Asymmetries on Scattered Neutron Spectra in ICF Implosions}
	\author{A. J. Crilly}
	\affiliation{Centre for Inertial Fusion Studies, The Blackett Laboratory, Imperial College, London SW7 2AZ, United Kingdom}
	\author{B. D. Appelbe}
	\affiliation{Centre for Inertial Fusion Studies, The Blackett Laboratory, Imperial College, London SW7 2AZ, United Kingdom}
	\author{O. M. Mannion}
	\affiliation{Laboratory for Laser Energetics, University of Rochester, Rochester, New York 14623, USA}
	\author{C. J. Forrest}
	\affiliation{Laboratory for Laser Energetics, University of Rochester, Rochester, New York 14623, USA}
	\author{J. P. Chittenden}
	\affiliation{Centre for Inertial Fusion Studies, The Blackett Laboratory, Imperial College, London SW7 2AZ, United Kingdom}
	\begin{abstract}
		Scattered neutron spectroscopy is a diagnostic technique commonly used to measure areal density in ICF experiments. Deleterious areal density asymmetries modify the shape of the scattered neutron spectrum. In this work a novel analysis is developed which can be used to fit the shape change. This allows scattered neutron spectroscopy to directly infer the amplitude and mode of the areal density asymmetries, with little sensitivity to confounding factors which affect other diagnostics for areal density. The model is tested on spectra produced by a neutron transport calculation with both isotropic and anisotropic primary fusion neutron sources. Multiple lines of sight are required to infer the areal density distribution over the whole sphere -- we investigate the error propagation and optimal detector arrangement associated with this inference. 
	\end{abstract}
	
	\maketitle
	
	\section{Introduction}
	
	Low mode drive asymmetries are proposed as a major degradation mechanism in Inertial Confinement Fusion (ICF) implosions. When mode 1 drive asymmetries are present, several experimental signatures are measured, these include: hotspot bulk flow velocity \cite{Hurricane2020,Hatarik2018,Mannion2018,Mannion2020}, anisotropy and differences in DT and DD inferred temperatures \cite{GatuJohnson2016} and asymmetric hotspot shape \cite{Ruby2016}. Hydrodynamic simulations suggest that the drive asymmetries, which can be used to explain these observations, will also cause corresponding areal density asymmetries which reduce confinement \cite{Spears2014,Chittenden2016,Clark2019,McGlinchey2018,Hurricane2020}. Recent work at the National Ignition Facility (NIF) has added to this evidence by exploring the relationship between hotspot velocity and areal density asymmetries measured using activation diagnostics \cite{Rinderknecht2020}. Activation diagnostics, while sensitive to areal density asymmetries, are also affected by hotspot velocity. Additional spectroscopic measurements are required to approximately remove this degeneracy \cite{Rinderknecht2018}. With scattered neutron images sophisticated tomographic techniques are required to extract the fuel density distribution \cite{Volegov2020}. Scattered neutron spectroscopy is another method commonly used to measure areal densities \cite{Johnson2012,Glebov2014}. It can be used at lower areal densities at which activation measurements are difficult, as is the case at OMEGA. While different areal densities can be measured on different detector lines of sight, the effect of areal density asymmetries on the scattered neutron spectrum has not been thoroughly investigated.
	
	In this paper we look to explore the effects of areal density asymmetries on the scattered neutron spectrum. The kinematic scattering relationships suggest that properties of the areal density asymmetries should be contained within a single measured spectrum. This will allow experimental spectroscopic measurements to leverage their high spectral resolution to directly infer more about the areal density distribution than currently possible.
	
	In \cref{section:scatteringgeometry} we will outline the geometry involved in neutron scattering. This will allow the areal density measured in neutron spectroscopy to be precisely defined. A model for a single scatter neutron spectrum from a DT ICF implosion will be described in \cref{section:spectralmodel}. This will be used to fit results from a neutron transport calculation in the following section. Finally in \cref{section:multipledetectors}, how to combine results from multiple lines of sight will be discussed. This will include an error analysis as well as a methodology to evaluate the optimal detector arrangement for general projection measurements, such as the hotspot velocity\cite{Hatarik2018}.
	
	\section{Scattering Geometry}\label{section:scatteringgeometry}
	
\begin{figure}[h]
		\centering
		\includegraphics*[width=0.85\columnwidth]{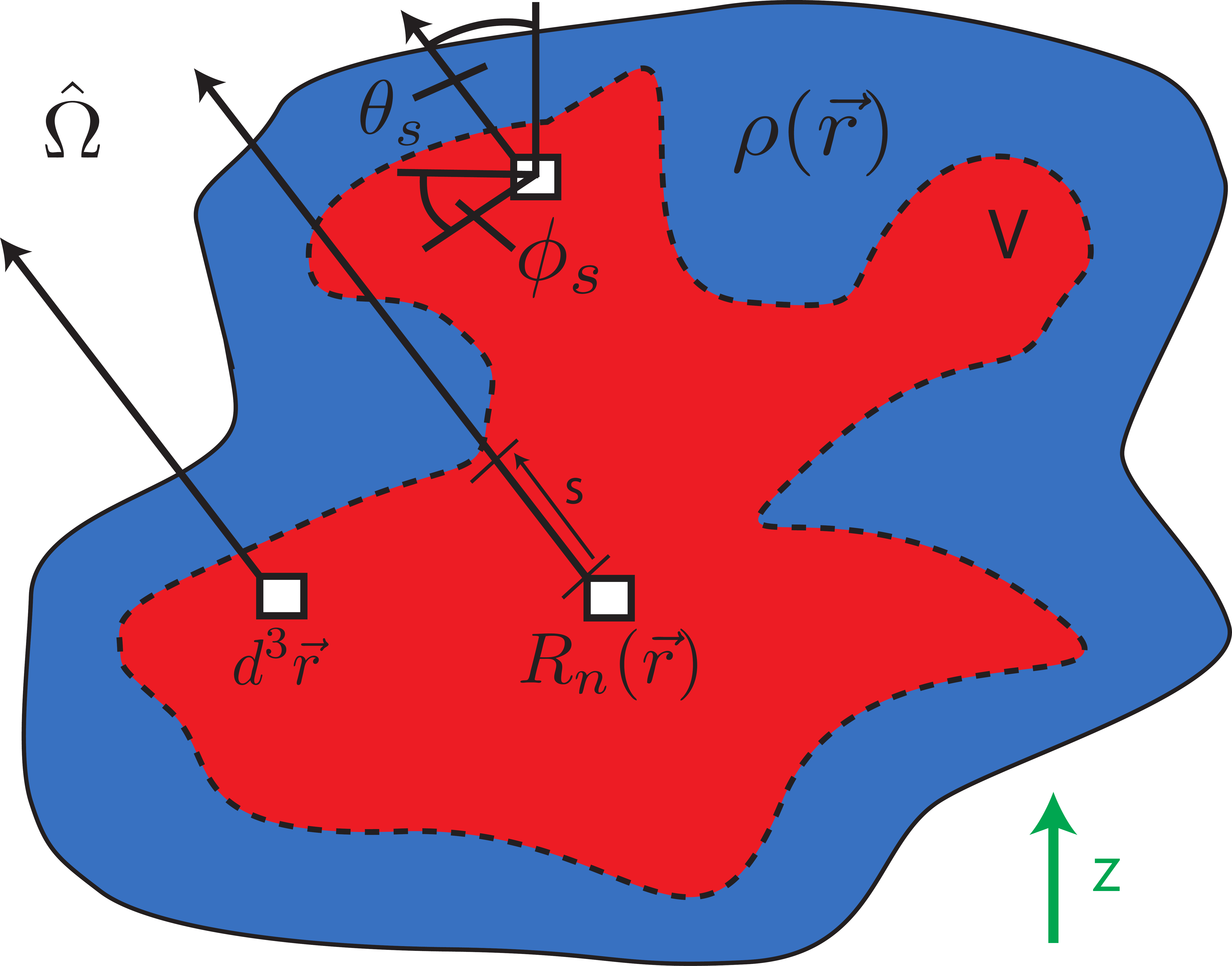}
		\caption{Diagram showing the geometry of the chord integrals given in \cref{eqn:rhoLdef,eqn:rhoLscdef,eqn:omegadef}. Line integrals of the density, $\rho$, along direction $\hat{\Omega}$ are taken from each point inside the central source region $V$. The chord integrals are weighted by the neutron production rate, $R_n$, at the source point.}
		\label{fig:chordintegraldiagram}
	\end{figure}

	The link between areal density asymmetries and scattered neutron spectral shape arises due to kinematic energy-angle relationships. These relationships are one-to-one for elastic processes, such as nD and nT scattering. From classical kinematics, in an elastic collision of a neutron with a stationary ion of mass $A m_n$, the incoming neutron energy, $E'$, and outcoming neutron energy, $E$, can be directly related to the scattering cosine, $\mu_s$:
	\begin{equation}\label{elastic}
	\frac{E}{E'} = \frac{\left(\mu_s+\sqrt{\mu_s^2+A^2-1}\right)^2}{\left(A+1 \right)^2} \ .
	\end{equation}
	Conversely, inelastic processes, such as n(D,2n)p and n(T,2n)D, produce a distribution of energies for a given scattering angle and vice versa. A feature which is shared amongst all nuclear interactions of interest is azimuthal symmetry. This allows us to define a `scattering cone' at a given scattering angle (or cosine) around which the differential cross section is only energy dependent. To understand the geometric interaction of scattering cones and areal density asymmetries we will consider the simplified case for which the birth spectra of primary fusion neutrons are isotropic. This allows separation of the spatial and spectral dimensions. The spatial effects on the neutron spectrum come through the areal density around the scattering cone. In ICF implosions, a central hotspot acts as an extended source of neutrons. Therefore we must consider the neutron-averaged line-integrated density, or $\langle \rho L \rangle$, rather than the more common $\rho R$. Working in the scattering cone geometry, we find $\langle \rho L \rangle_{\mathrm{s.c.}}$ along chords at a fixed scattering angle (with respect to our detector). For simplicity, we will consider the detector to lie along the $z$-axis, i.e. $\theta = 0$. We then consider the areal density seen by a beam of neutrons born at $\vec{r}$, with initial direction, $\hat{\Omega}$, before they scatter into the detector line of sight:
	\begin{equation}
		\rho L =\int_0^{\infty} \rho (\vec{r}+s\hat{\Omega}) ds \ .
	\end{equation}
	To neutron-average we must sum over the whole reacting volume. The $\langle \rho L \rangle$ and $\langle \rho L \rangle_{\mathrm{s.c.}}$ are then defined as follows:
	\begin{align}
		\langle \rho L \rangle(\theta_s,\phi_s) &= \frac{1}{Y_n}\int_V d^3\vec{r}\int ds \rho (\vec{r}+s\hat{\Omega}) R_n(\vec{r}) \ , \label{eqn:rhoLdef} \\
		\langle \rho L \rangle_{\mathrm{s.c.}}(\theta_s) &= \int \langle \rho L \rangle(\theta_s,\phi_s) d\phi_s \ , \label{eqn:rhoLscdef} \\
		\hat{\Omega} &\equiv \left[\sin(\theta_s)\cos(\phi_s),\sin(\theta_s)\sin(\phi_s),\cos(\theta_s)\right] \label{eqn:omegadef}
	\end{align}
	where $Y_n$ is the fusion neutron yield, $\rho$ is the mass density and $R_n$ is the neutron production rate. The geometry of these chord integrals is shown in \cref{fig:chordintegraldiagram}.
	For a point source, $\langle \rho L \rangle$ will coincide with the $\rho R$. However, this is not a good approximation for a typical ICF implosion. With perfect spherical symmetry, $\langle \rho L \rangle$ is proportional to $\rho R$ with a coefficient which depends on hydrodynamic profiles \cite{Johnson2012,Frenje2013}. In asymmetric conditions $\langle \rho L \rangle$ is well-defined and measurable, while $\rho R$ is not.
	
	As $\langle \rho L \rangle$ is defined over the surface of a sphere it is natural to expand it in terms of spherical harmonics, $Y_l^m$:
	
	\begin{equation}
		\langle \rho L \rangle(\theta_s,\phi_s) = \sum_{l=0}^{\infty} \sum_{m=-l}^{m=+l} \langle \rho L \rangle_l^m Y_l^m(\theta_s,\phi_s) \ .
	\end{equation}
	
	Since the scattering cone integrates over azimuthal angle, $\phi_s$, it is simple to show that:
	\begin{equation}
	\langle \rho L \rangle_{\mathrm{s.c.}}(\theta_s) = \sum_{l=0}^{\infty} \langle \rho L \rangle_l P_l(\mu_s) \ ,
	\end{equation}
	where $P_l$ are the Legendre polynomials and $\mu_s$ is the scattering cosine ($\cos(\theta_s)$). Thus, a Legendre expansion in $\langle \rho L \rangle_{\mathrm{s.c.}}$ is general as the spherical harmonic expansion of $\langle \rho L \rangle$ is complete. The areal density Legendre coefficients, $\langle \rho L \rangle_l$, are directly proportional to the $m=0$ spherical harmonic coefficients. Modes with $m \neq 0$ vanish during azimuthal integration so do not affect $\langle \rho L \rangle_{\mathrm{s.c.}}$. Therefore scattered spectra are blind to $m \neq 0$ modes. Due to the extended nature of the hotspot neutron source, $\langle \rho L \rangle$ is likely to be dominated by low modes \cite{Crilly2018}.
	
	Note that these expansions are defined in reference to a coordinate system centred on the detector line of sight. The coefficients of order $l$ along one line of sight are a linear combination of the coefficients, also of order $l$, on a different line of sight.
	
	For mode $l=1$ areal density asymmetries, this is particularly simple:
	\begin{equation}
	\langle \rho L \rangle_{\mathrm{s.c.}}^{l=1}(\mu_s) = \langle \rho L \rangle_0 + \left[\vec{\langle \rho L \rangle}_1 \cdot \hat{\Omega}_{\mathrm{det}} \right] \mu_s \ , \label{eqn:mode1rhoL}
	\end{equation}
	where $\hat{\Omega}_{\mathrm{det}}$ is the detector line of sight and $\vec{\langle \rho L \rangle}_1$ is a vector which points along the axis of the mode 1 and has a magnitude given by the half the peak to trough areal density asymmetry. The addition of an isotropic and vector component resembles the formula for the DT primary centroid shift \cite{Munro2016}. Therefore a similar 4 detector set up \cite{Hatarik2018,Mannion2018,Mannion2020} is required in order to constrain the mode 1 areal density asymmetry using scattered spectra. For every higher mode an additional $2l + 1$ detectors are required in order to back out the physical $\langle \rho L \rangle_l^m$ coefficients.
	
	The energy spectrum of the scattered neutrons depends on the spectrum of primary fusion neutrons and the areal density distribution. The nuclear interaction differential cross sections provide the link between these spectral and spatial dependencies. Given an birth neutron energy spectrum, $Q_b(E')$, the scattered neutron spectrum is found as follows:
	
	\begin{align}
		\frac{d^2N}{d\mu_sdE} &= \frac{\langle \rho L \rangle_{\mathrm{s.c.}}(\mu_s)}{\bar{m}} \int \frac{d^2 \sigma}{d\mu_s dE} Q_b(E') dE' \ , \label{eqn:scatteringconedNdE}\\
		\frac{dN}{dE} &= \int \frac{d^2N}{d\mu_sdE} d\mu_s \ , \label{eqn:dNdE}
	\end{align}
	where $\bar{m}$ is the average ion mass and the double differential cross section is the first term inside the integrand in \cref{eqn:scatteringconedNdE}. The $dN/dE$ can be measured via time-of-flight or magnetic recoil spectrometers. The primary neutron spectrum is also measured and this can be used to separately constrain the birth spectrum, this leaves the areal density distribution as the single unknown. These equations will stand as the basis for a model to describe the singly scattered neutron spectrum in asymmetric ICF implosions.

	\section{Spectral Model}\label{section:spectralmodel}
	
	Using the scattering cone integration, one can construct full singly scattered neutron spectrum model for asymmetric ICF implosions. The single scatter approximation limits the scope of this model to areal densities $\lesssim$ 200 mg/cm$^2$. In this work we will focus on this areal density regime. The single scatter approximation also implies that the level of attenuation is low since the total cross section is dominated by scattering rather than absorption. Multiple scattering becomes increasingly important at higher areal densities. One can extend the scattering cone method to multiple scattering but many of the approximations made are invalid. More accurate models for higher areal densities will be the subject of future work. Working with higher energy scattered neutrons can also be used to reduce the impact of multiple scattering \cite{Crilly2018}. At these higher energies, both the scattering cross section and the number of multiply scattered neutrons are lower.
	
	The set of reactions we will consider are: D(T,n)$\alpha$, D(D,n)$^3$He, T(T,2n)$\alpha$, nT elastic, nD elastic, n(D,2n)p and n(T,2n)D. Cross section data for these reactions were taken from the ENDF\cite{ENDF} and CENDL\cite{CENDL} nuclear data libraries. The Bosch-Hale\cite{Bosch1992} DT and DD fusion reactivities and the Appelbe\cite{Appelbe2016} calculation of the TT temperature dependent spectral shape were used. The DT and DD primary moments were calculated using relativistically correct expressions derived by Ballabio\cite{Ballabio1998}. For stationary target ions, the differential cross sections can simply be taken from nuclear data libraries, as will be done in this model. The effects of scattering ion velocities can be included for elastic processes using the method outlined in Crilly \textit{et al.} \cite{Crilly2020}. These ion velocity effects are particularly important for the spectral shape of the backscatter edges.
	
\begin{figure}[h]
		\centering
		\includegraphics*[width=1.0\columnwidth]{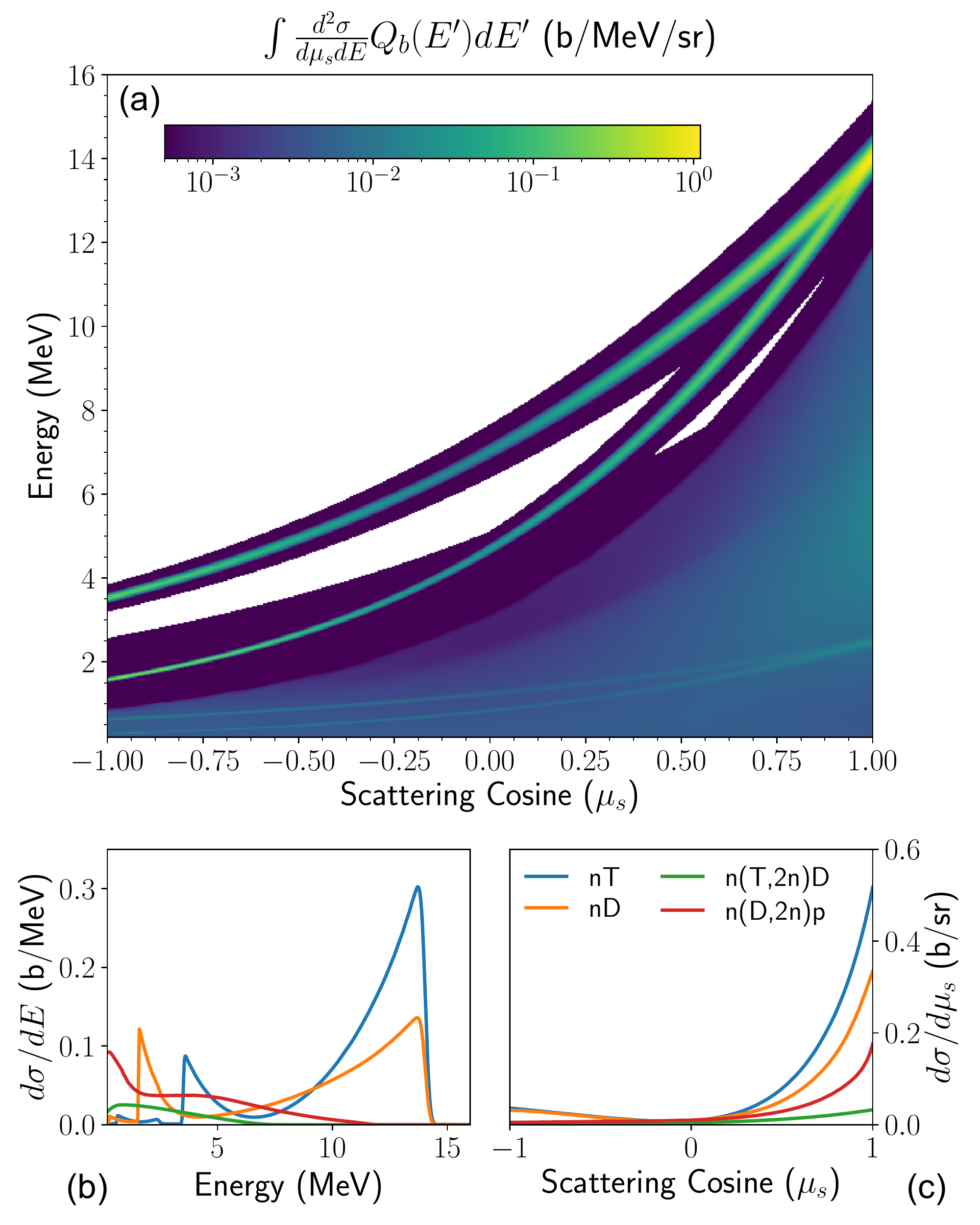}
		\caption{(a) The DT double differential cross section integrated over all primary fusion sources. The two high intensity lines correspond to nD and nT elastic scattering of DT fusion neutrons. The white areas of the plot are kinematically forbidden. The spectral shapes and yield ratios of the primary neutron sources were calculated assuming $\langle T_i \rangle_{\mathrm{DT}}$ = 4 keV and 50/50 DT isotopic composition. (b) A plot of the differential cross section with respect to outgoing energy for each scattering component. (c) A plot of the differential cross section with respect to scattering cosine for each scattering component. These single differential cross sections are marginalisations of the double differential cross section.}
		\label{fig:dndedmus_model}
	\end{figure}

	First we will look at the form of the scattering cone neutron spectrum as given in \cref{eqn:scatteringconedNdE}. The double differential cross section term can be calculated independently of the areal density distribution. The birth neutron spectrum of the primary fusion reactions can be calculated using the models listed above. For a stationary uniform plasma the spectral shapes and yields of the primaries only dependent on a single parameter, the ion temperature. More complex source conditions will alter the birth spectrum but the model will aim to describe the birth spectra using only a single temperature, taken as the burn averaged DT ion temperature, $\langle T_i \rangle_{\mathrm{DT}}$. \Cref{fig:dndedmus_model} shows the differential cross section terms for a 4 keV ion temperature case.

\begin{figure}[h]
	\centering
	\includegraphics*[width=0.99\columnwidth]{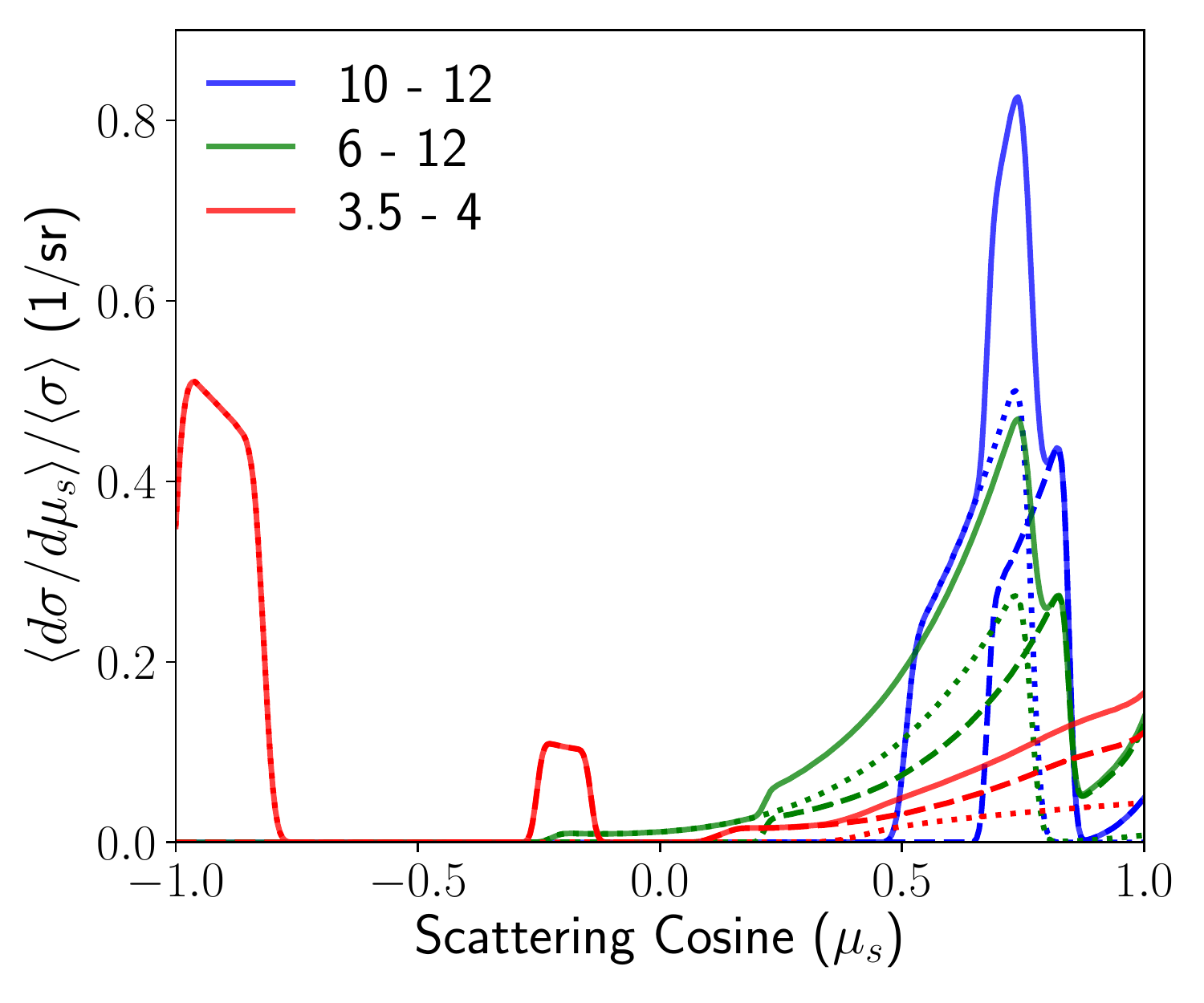}
	\caption{The angular weighting function for various down scatter ratio ranges in MeV. The solid lines show the total, the dashed lines show the D contribution (from all scattering processes) and the dotted lines show the T contribution.}
	\label{fig:DSR_weighting}
	\end{figure}

	While limiting the energy range of a spectral measurement might be dominated by a particular angular range, there are still contributions from other scattering angles. This becomes increasingly important at lower energies where many scattering sources have similar amplitude. Areal density measurements found using a down-scatter ratio (DSR) technique will therefore find the following average measurement of the areal density distribution:
	\begin{align}
	\langle \rho L \rangle_{\mathrm{DSR}} &= \frac{\bar{m}}{\langle \sigma \rangle} \frac{\int_{\mathrm{DSR}}\frac{dN}{dE}dE}{\int_{13}^{15}\frac{dN}{dE}dE} \ , \\
	&\approx  \frac{\int d\mu_s \langle \rho L \rangle_{\mathrm{s.c.}}(\mu_s) \int_{\mathrm{DSR}} dE \left.\frac{d^2 \sigma}{d\mu_s dE}\right|_{\mathrm{DT}}}{ \int d\mu_s \int_{\mathrm{DSR}} dE \left.\frac{d^2 \sigma}{d\mu_s dE}\right|_{\mathrm{DT}}} \ , \label{eqn:DSRweighting}
	\end{align}
	where the notation on the double differential cross sections indicates that they have been integrated over the DT peak only. Note we convert the DSR to a $\langle \rho L \rangle$, not a $\rho R$. The conversion to $\rho R$ is sensitive to hydrodynamic profiles and therefore varies between implosions. Estimates for this geometry effect exist in the literature\cite{Johnson2012,Frenje2013}. \Cref{fig:DSR_weighting} shows the angular weighting function for various DSR ranges.
	Inferring areal density asymmetries using DSR techniques from multiple lines of sight will therefore have to account for this averaging effect as well as the line of sight projection effect of the $\langle \rho L \rangle$ distribution. Therefore, it may be advantageous to perform a fit to scattered spectrum over a large energy range to infer the $\langle \rho L \rangle_l$ coefficients directly. Fitting the spectrum also has the additional advantage that a single line of sight measurement can be used to quantify areal density asymmetries whereas DSRs require multiple lines of sight to do this. This will be investigated further in the following sections of this work.
	
\begin{figure}[h]
		\centering
		\includegraphics*[width=0.99\columnwidth]{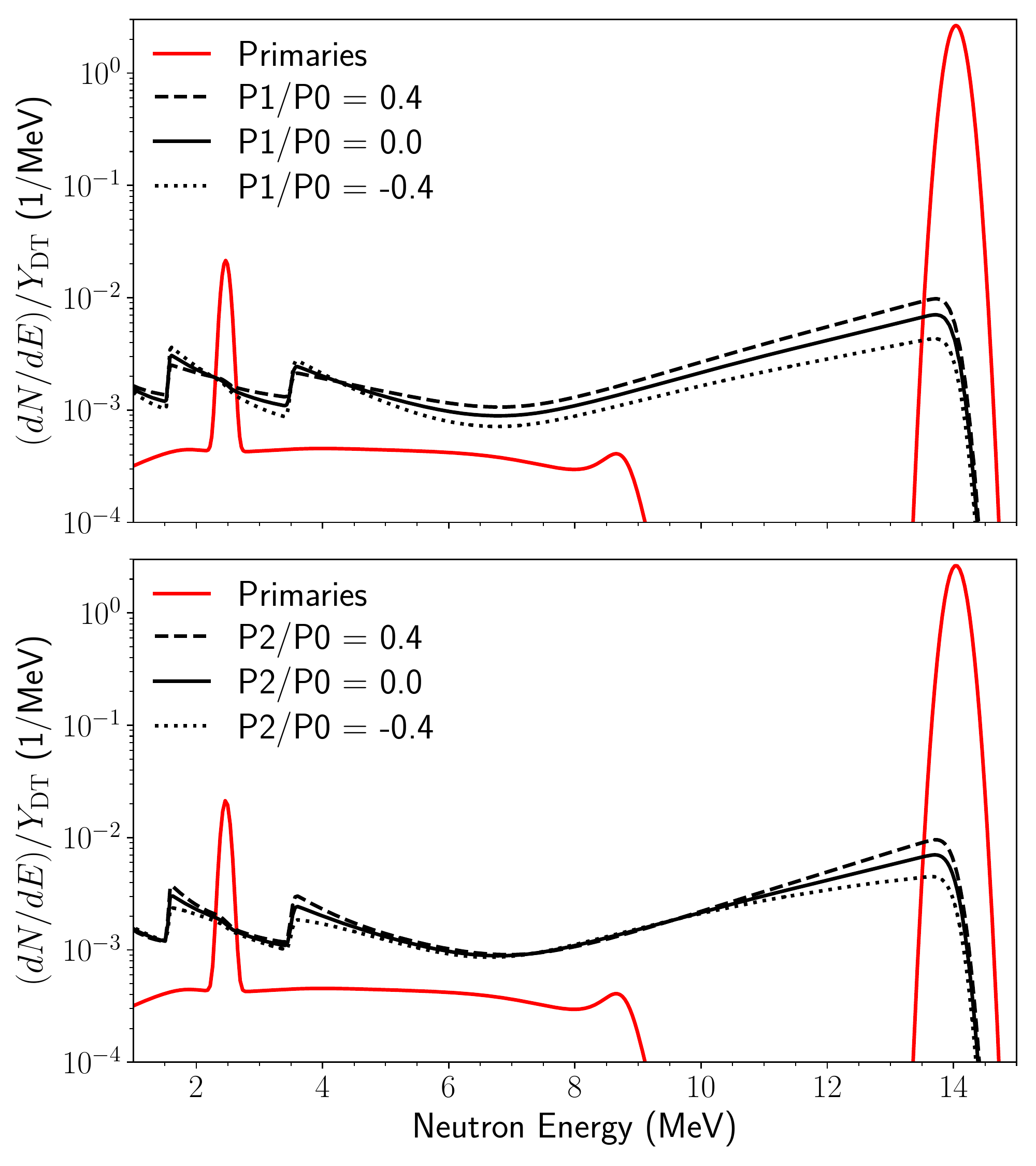}
		\caption{The full primary and scattered neutron spectrum for various $l=1$ (top) and $l=2$ (bottom) areal density asymmetry amplitudes. The primary spectrum shapes correspond to a 4 keV 50/50 DT plasma. The $\langle \rho L \rangle_0$ was taken to be 150 mg/cm$^2$ for both cases.}
		\label{fig:dnde_model}
	\end{figure}

	Given the scattering cone neutron spectrum, one can compute the resultant total neutron spectrum by integration over scattering cosine (as shown in \cref{eqn:dNdE}). \Cref{fig:dnde_model} shows the spectrum for various $l=1$ and $l=2$ areal density asymmetry amplitudes.
	\begin{figure}[h]
		\centering
		\includegraphics*[width=0.75\columnwidth]{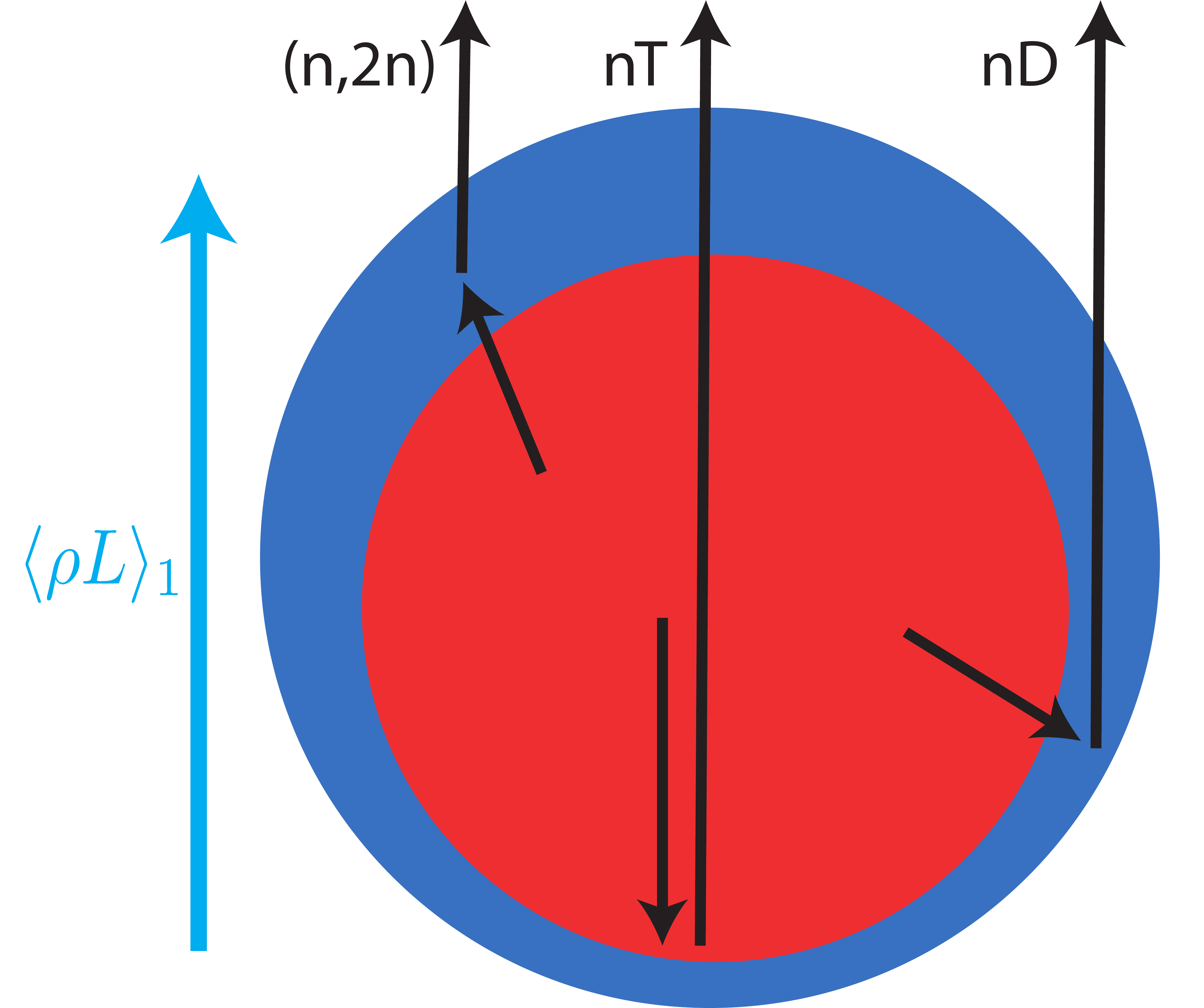}
		\caption{Diagram showing some example scattering trajectories for different nuclear interactions which produce outgoing neutrons at $\approx$ 3.5 MeV. The hotspot and shell configuration, shown by the red and blue regions respectively, results in a mode 1 areal density asymmetry in the direction indicated in the diagram.}
		\label{fig:Mode1Diagram}
	\end{figure}
	Deviations from the symmetric case are notable and different behaviour is observed for $l=1$ and $l=2$. A simple interpretation based on the elastic energy-angle relationship in \cref{elastic} fails to predict some of the behaviour. For example, for $l=1$ one might expect a positive mode 1 to be lower amplitude compared to the symmetric case at the backscatter regions. This is not the case due to the (n,2n) reactions which have a very forwarded peaked differential cross section (c.f. \cref{fig:dndedmus_model}). Their contribution to the elastic backscatter regions samples areal density from the forward direction -- opposite to that sampled by the elastic interactions (c.f. the 3.5 - 4 MeV result in \cref{fig:DSR_weighting}). A diagrammatic representation of this effect is given in \cref{fig:Mode1Diagram}. This demonstrates the need for the model outlined in this section in order to understand the effects of areal density asymmetries on neutron spectra.

	\section{Neutron Transport Comparison}
	
\begin{figure}[h]
		\centering
		\includegraphics*[width=0.99\columnwidth]{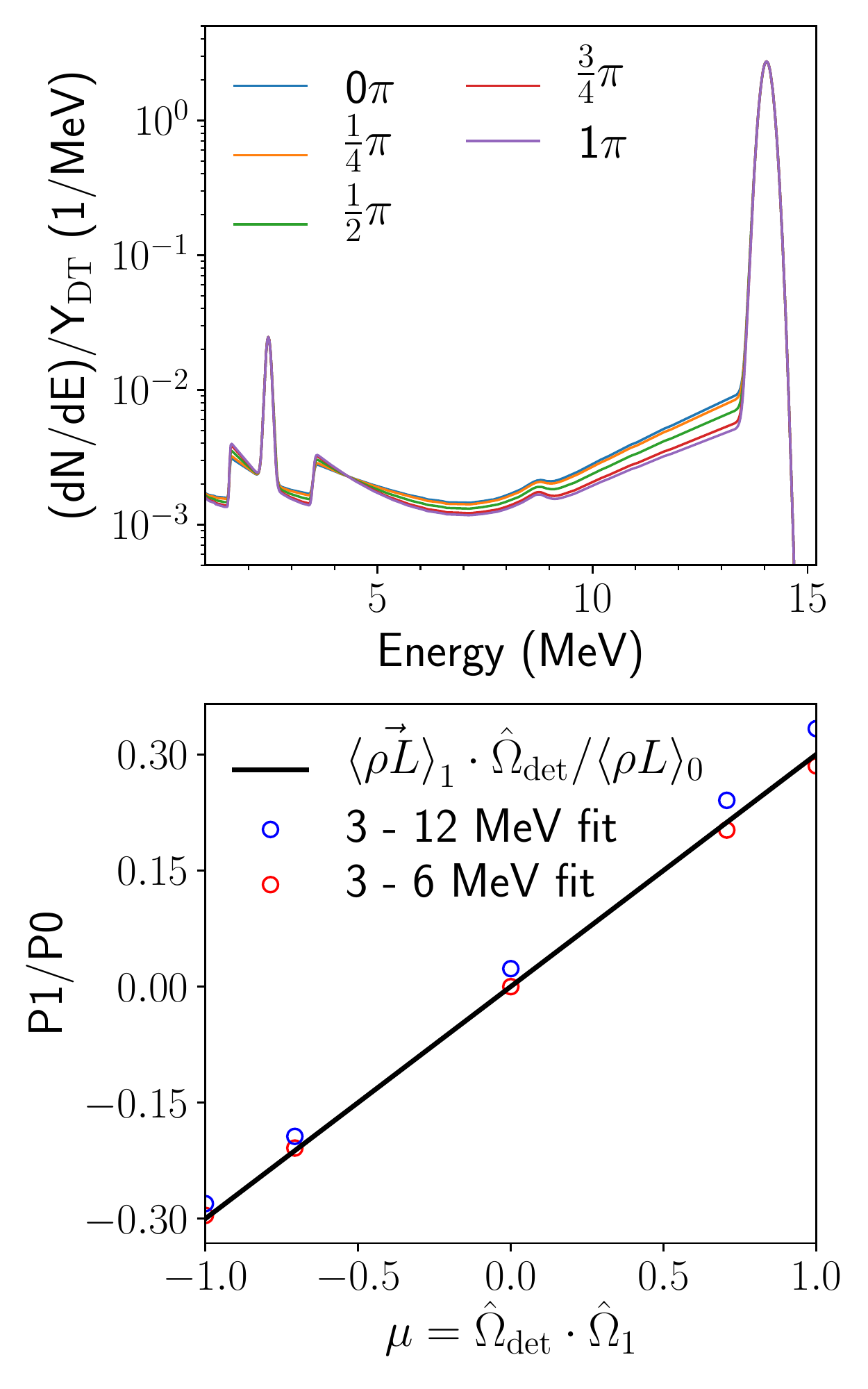}
		\caption{(Top) Plot showing the neutron spectra from a mode 1 ice block model at various angles to the asymmetry axis. (Bottom) Results from fitting the spectra in two different energy ranges. These fits were used to infer the magnitude of mode 1 areal density asymmetry. The theoretical result from \cref{eqn:mode1rhoL} is shown by the solid black line as a function of the cosine between detector and mode 1 direction, $\mu$.}
		\label{fig:iceblockstationary}
	\end{figure}

	The model laid out in \cref{section:spectralmodel} will be compared to results from a more complete neutron transport description of an ICF implosion. The inverse ray trace method described in Crilly \textit{et al.} \cite{Crilly2018} allows the calculation of neutron spectra given 3D grids of hydrodynamic conditions. As a suitable test problem, we consider an ice block model for a mode 1 asymmetry where a stationary spherical hotspot is displaced within a dense cold fuel layer, as in \cref{fig:Mode1Diagram}. The hotspot was chosen to have a parabolic temperature profile to test the spectral model's reliance on a single average temperature. The ice block configuration was then chosen such that $\langle T_i \rangle_{DT}$ = 3.9 keV, $\langle \rho L \rangle_0$ = 150 mg/cm$^2$ and $\langle \rho L \rangle_1/\langle \rho L \rangle_0$ = 30\%. The detector line of sight was then rotated with respect to the mode 1 axis to explore the relationship derived in \cref{eqn:mode1rhoL}. Neutron spectra were calculated for these different detectors using the inverse ray trace method. These spectra were fitted using the model outlined in \cref{section:spectralmodel} to obtain $\langle \rho L \rangle_0$ and $\langle \rho L \rangle_1$ for each detector. The results of this analysis are shown in \cref{fig:iceblockstationary}. 
	Every fit retrieved a value of $\langle \rho L \rangle_0$ within 4 mg/cm$^2$ of the true value of 150 mg/cm$^2$. This error originates in the approximations of the model, namely single temperature primary spectra and no differential attenuation. For $\langle \rho L \rangle_1$, it was found that a fit to the spectrum between 3-6 MeV produced the best results. This range of energies is sensitive to all scattering angles (c.f. \cref{fig:dndedmus_model}) but also narrow enough that the effect of differential attenuation of the scattered neutrons is small. Narrow high energy ranges are more robust to attenuation effects but have lower angular sensitivity.
	
	It is common for the primary spectra to exhibit anisotropy due to bulk fluid flow of the fusing plasma. For a mode 1 drive asymmetry, the hotspot velocity is aligned with the areal density asymmetry\cite{Spears2014,Chittenden2016,Rinderknecht2020,Hurricane2020}. According to the experimental data analysis by Rinderknecht \textit{et al.} \cite{Rinderknecht2020}, the following linear relationship between mode 1 areal density asymmetry and hotspot velocity matches experimental data:
	\begin{equation}
	\frac{\langle \rho L \rangle_1}{\langle \rho L \rangle_0} = 0.39 \frac{v_{\mathrm{HS}}}{100 \mathrm{km/s}} \ .
	\end{equation}
	Thus the 30\% asymmetry presented here corresponds to a hotspot velocity $\sim$ 77 km/s. To see the effect of a spectral anisotropy of this magnitude, a uniform hotspot velocity (of 77 km/s) is introduced to the ice block model. For the fitting method, the spectral anisotropy was included through the birth spectrum term, $Q_b(E',\mu_s,v_{\mathrm{HS}})$, in \cref{eqn:scatteringconedNdE} with the assumption that the mode 1 areal density asymmetry and hotspot flow velocity were aligned. It was found that the same results as shown for the stationary hotspot case were obtained when the hotspot flow velocity was properly accounted for. If the anisotropy of the birth spectrum was not included then an additional absolute $\sim 1 \%$ error was introduced to the inferred $\langle \rho L \rangle_1/\langle \rho L \rangle_0$.
	
\begin{figure}[h]
		\centering
		\includegraphics*[width=0.99\columnwidth]{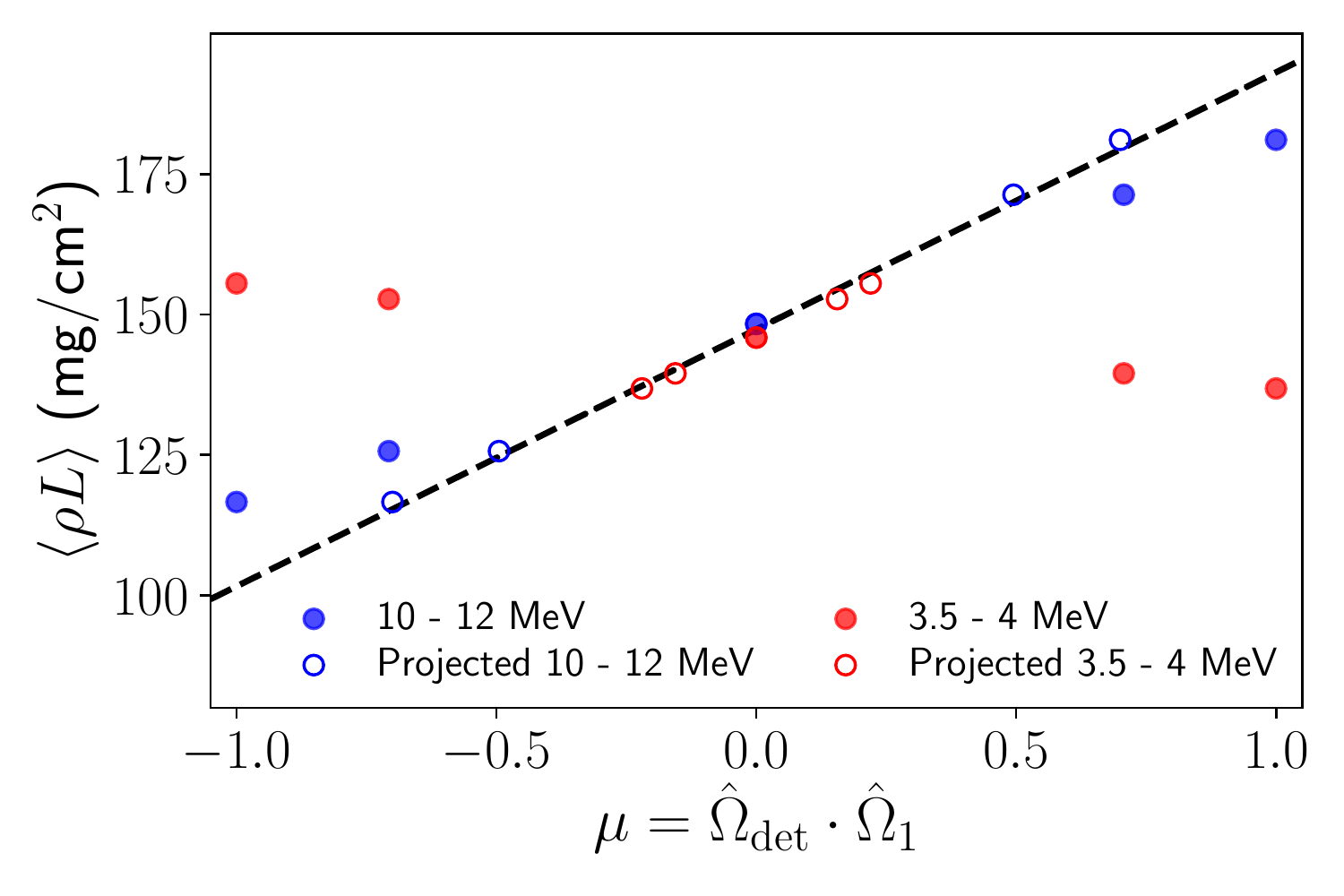}
		\caption{Down scatter ratio inferred areal density (using \cref{eqn:rhoLDSR}) as a function of the angle between the detector line of sight and the mode 1 axis. The results are projected using the average scattering cosine as given in \cref{eqn:avgDSRcos}, shown as hollow circles. The black dashed shows a linear fit to the projected areal densities.}
		\label{fig:iceblockDSR}
	\end{figure}

	Finally, an areal density inference based on DSRs was performed. For a pure mode 1, the $\langle \rho L \rangle_{\mathrm{DSR}}$ as defined in \cref{eqn:DSRweighting} can be simplified:
	\begin{align}\label{eqn:rhoLDSR}
	\langle \rho L \rangle_{\mathrm{DSR}} &= \langle \rho L \rangle_0 + \langle \rho L \rangle_1 \langle \mu_s \rangle_{\mathrm{DSR}} \ , \\
	\langle \mu_s \rangle_{\mathrm{DSR}} &\equiv \int \mu_s \frac{1}{\langle \sigma \rangle} \left\langle \frac{d\sigma}{d\mu_s} \right\rangle d\mu_s \ , \label{eqn:avgDSRcos}
	\end{align}
	where $\langle \mu_s \rangle_{\mathrm{DSR}}$ is the average scattering cosine being sampled in the DSR range. The 10 -- 12 MeV and 3.5 -- 4 MeV ranges have $\langle \mu_s \rangle_{\mathrm{DSR}}$ values of 0.70 and -0.22 respectively (based on the results of \cref{fig:DSR_weighting}). Unlike the fitting method, values of $\langle \rho L \rangle_0$ and $\langle \rho L \rangle_1$ cannot be inferred separately on a single line of sight when using DSRs. \Cref{fig:iceblockDSR} shows the inferred $\langle \rho L \rangle_{\mathrm{DSR}}$ for the ice block spectra presented in \cref{fig:iceblockstationary}. 
	Without taking into account the average scattering cosine, $\langle \mu_s \rangle_{\mathrm{DSR}}$, the DSRs as a function of viewing angle is a poor representation of the underlying areal density asymmetry. This is due to the fact that the areal density probed by the DSR measurement is not along the detector line of sight. Because of this, interpreting the $\langle \rho L \rangle$ through measured DSRs requires additional care. If the underlying areal density distribution is assumed to be mode 1, a linear fit can be performed once the measurements have been projected using the appropriate angle, $\langle \mu_s \rangle_{\mathrm{DSR}}$. For higher modes \cref{eqn:rhoLDSR} must be expanded to higher order introducing higher moments of $\mu_s$, complicating further the relationship between DSR and underlying physical areal density asymmetries. The results of a mode 1 fit to the data in \cref{fig:iceblockDSR} are values of $\langle \rho L \rangle_0$ = 148 mg/cm$^2$ and $\langle \rho L \rangle_1/\langle \rho L \rangle_0$ = 30\%, within a few percent of the true values. Here we have used the fact we know the direction of the mode 1 \textit{a priori} to analyse the data, in the next section we will consider the general case of a completely unknown $\vec{\langle \rho L \rangle}_1$.
	
	\section{Multiple Detectors and Error Analysis}\label{section:multipledetectors}
	As shown in \cref{section:scatteringgeometry}, a single measurement of the scattered neutron spectrum contains incomplete information about the physical $\langle \rho L \rangle(\theta,\phi)$ distribution. Multiple measurements are required in order to infer the underlying spherical harmonic moments. The inference of hotspot flows from primary neutron spectra requires similar analysis \cite{Hatarik2018,Mannion2020}. In this section we will consider how to infer the mode 1 areal density asymmetry vector, $\vec{\langle \rho L \rangle}_1$, and the error analysis. We will approach this as a general projection measurement first. We will consider the case where we have $n_d$ detectors and $n_d \geq n_p$, where $n_p$ is the number of physical parameters we are inferring. The measurements can then be related to the underlying physical parameters with the following system of linear equations:
	\begin{align}
		\vec{m} &= \mathbf{A}\cdot\vec{p} 
	\end{align}
	where $\vec{m}$ is the length $n_d$ vector of measurements, $\mathbf{A}$ is the projection $n_p\times n_d$ matrix and $\vec{p}$ is the length $n_p$ vector of physical parameters. The normal equation can be used to find the least squares distance between measurement and inferred physical parameters. This leads to the definition of the pseudo-inverse, $\mathbf{A}^+$:
	\begin{align}
	\mathbf{A}^T\vec{m} &= \mathbf{A}^T\mathbf{A}\cdot\vec{p} \ , \\
	\vec{p} &= \mathbf{A}^+\vec{m} \ , \\
	\mathbf{A}^+ &\equiv (\mathbf{A}^T\mathbf{A})^{-1}\mathbf{A}^T \ .
	\end{align}
	We are also interested in how error propagates from measurement to inferred values. By considering the covariance matrix of $\vec{p}$, which is denoted $\Sigma_p$, one can find a detector arrangement that minimises this error:
	\begin{align}\label{eqn:covarpdef}
	\Sigma_p &= \mathbf{A}^+\Sigma_m\left(\mathbf{A}^+\right)^T \ .
	\end{align}
	In order to precede with the analysis we will assume that the detectors are independent and have the same error, $\sigma_m$. Then, the covariance matrix of the measurements $\Sigma_m = \sigma_m^2 \mathbf{I}$. Using the properties of symmetric matrices for $\mathbf{A}^T\mathbf{A}$, this simplifies \cref{eqn:covarpdef}:
	\begin{align}
	\Sigma_p &= \sigma_m^2\mathbf{A}^+\left(\mathbf{A}^+\right)^T = \sigma_m^2(\mathbf{A}^T\mathbf{A})^{-1}\mathbf{A}^T\mathbf{A}\left[(\mathbf{A}^T\mathbf{A})^{-1}\right]^T \ , \nonumber \\
	&= \sigma_m^2(\mathbf{A}^T\mathbf{A})^{-1} \ .
	\end{align}
	The determinant of $\Sigma_p$ is a measure of the total error in $\vec{p}$. Therefore the determinant of $\mathbf{A}^T\mathbf{A}$ must be maximised to minimise this error.
	
	We will investigate the optimal detector arrangement for multiple DSR measurements as a worked example of this methodology. The projection matrix, $\mathbf{A}$, for the DSR method can be found by combining \cref{eqn:rhoLDSR,eqn:mode1rhoL}. 
	\begin{align}
	\begin{bmatrix}
	\langle \rho L \rangle_{\mathrm{DSR}}^{(1)}  \\
	\vdots \\
	\langle \rho L \rangle_{\mathrm{DSR}}^{(n_d)} 
	\end{bmatrix} &= 
	\begin{bmatrix}
	1 & u^{(1)}_x & u^{(1)}_y & u^{(1)}_z \\
	\vdots & \vdots & \vdots & \vdots \\
	1 & u^{(n_d)}_x & u^{(n_d)}_y & u^{(n_d)}_z
	\end{bmatrix}\cdot
	\begin{bmatrix}
	\langle \rho L \rangle_0     \\
	\langle \rho L \rangle_{1,x} \\
	\langle \rho L \rangle_{1,y} \\
	\langle \rho L \rangle_{1,z}
	\end{bmatrix}. \\
	\vec{u}^{(i)} &\equiv \langle \mu_s \rangle_{\mathrm{DSR}}^{(i)}\hat{\Omega}^{(i)} \ ,
	\end{align}
	where the bracketed superscripts denote the detector index. If one sets $\langle \mu_s \rangle_{\mathrm{DSR}}^{(i)} = 1$, then this projection matrix also applies for the measurement of vector hotspot velocity and isotropic Gamow shifts for primary DT and DD neutron spectra \cite{Hatarik2018,Mannion2020}. Evaluating $\mathbf{A}^T\mathbf{A}$ gives the following 4x4 matrix:
	\begin{align}
	\mathbf{A}^T\mathbf{A} &= n_d \begin{bmatrix}
	1 & \langle u_x\rangle & \langle u_y\rangle & \langle u_z\rangle \\
	\langle u_x\rangle & \langle u_x^2\rangle & \langle u_xu_y\rangle & \langle u_xu_z\rangle \\
	\langle u_y\rangle & \langle u_xu_y\rangle & \langle u_y^2\rangle & \langle u_yu_z\rangle \\
	\langle u_z\rangle & \langle u_xu_z\rangle & \langle u_yu_z\rangle & \langle u_z^2\rangle 
	\end{bmatrix} \\
	\langle u_j\rangle &= \frac{1}{n_d}\sum_{i=1}^{n_d} u_j^{(i)} \ .
	\end{align}
	In order for the inferred quantities to have no covariances then this matrix must be diagonal. Setting the off-diagonal components to zero will sets requirements on the detector lines of sight. As discussed earlier, the minimal error detector arrangement is the one that maximises the determinant:
	\begin{align}
	\mbox{det}\left(\mathbf{A}^T\mathbf{A}\right) &= n_d^4 \ \mbox{det}\begin{bmatrix}
	1 & \langle u_x\rangle & \langle u_y\rangle & \langle u_z\rangle \\
	\langle u_x\rangle & \langle u_x^2\rangle & \langle u_xu_y\rangle & \langle u_xu_z\rangle \\
	\langle u_y\rangle & \langle u_xu_y\rangle & \langle u_y^2\rangle & \langle u_yu_z\rangle \\
	\langle u_z\rangle & \langle u_xu_z\rangle & \langle u_yu_z\rangle & \langle u_z^2\rangle 
	\end{bmatrix} \nonumber \ , \\
	&= n_d^4 \ \mbox{det}\left( \langle \vec{u}\vec{u}^T\rangle - \langle \vec{u}\rangle \langle\vec{u}^T\rangle \right) \ , \\
	&= n_d^4 \ \mbox{det}\left( \Sigma_{u}\right) \ ,
	\end{align}
	where $\Sigma_{u}$ is the 3x3 covariance matrix of detector projection vectors, $\vec{u}$. We will investigate the case where every detector uses the same DSR range such that $\langle \mu_s \rangle_{\mathrm{DSR}}^{(i)} = \langle \mu_s \rangle_{\mathrm{DSR}}$. The angular factors can then be factored out:
	\begin{align}
	\mbox{det}\left(\mathbf{A}^T\mathbf{A}\right) &= n_d^4 \langle \mu_s \rangle_{\mathrm{DSR}}^6 \ \mbox{det}\left( \Sigma_{\Omega}\right) \ ,
	\end{align}
	leaving only dependence on the (unit length) line of sight vectors, $\hat{\Omega}^{(i)}$. Therefore, we can now use statistical results for the analysis of spherical data. The normalised orientation matrix, $\mathbf{T} = \langle \hat{\Omega}\hat{\Omega}^T\rangle$, has positive eigenvalues, $\tau_i$, that sum to unity \cite{Fisher1993}. Its determinant is then maximised when all these eigenvalues are equal. This describes an uniform isotropic distribution for which:
	\begin{align}
		\tau_1 &= \tau_2 = \tau_3 = \frac{1}{3} \ \mbox{and} \ \langle \hat{\Omega}\rangle = \vec{0} \ ,
	\end{align}
	leading to: 
	\begin{align}
		\mbox{det}\left( \Sigma_{\Omega}\right) = \mbox{det}\left(\mathbf{T}\right) &= \frac{1}{27} \ .
	\end{align}
	These requirements also lead to a diagonal $\mathbf{A}^T\mathbf{A}$ with values ($n_d$,$n_d$/3,$n_d$/3,$n_d$/3). For $n_d = 4$, the optimum is found for a tetrahedral arrangement. The solution to this problem is analogous to arranging masses on the unit sphere such that the centre of gravity is at the origin and all principal moments of inertia are equal. This second constraint is satisfied if there are two axes with n-fold (n $\geq 3$) rotational symmetry \cite{Aravind1992}.
	
	Using the analysis outlined in this section, the optimal line of sight for a new detector given the current ($n_d = 4$) NIF nToF arrangement\cite{Hatarik2018} would be along a ($\theta$--$\phi$) of (77--68). Similarly at OMEGA, for the current nToF arrangement\cite{Mannion2020} ($n_d = 5$ unique lines of sight), the next optimal line of sight is along (135--301).
	
	The analysis is more involved if multiple different DSR ranges are used as the angle terms cannot be factored out. The optimisation problem must then be solved separately for any given combination of DSR ranges. 

	\section{Conclusions}
	
	Scattered neutron spectroscopy contains valuable information on areal density asymmetries in ICF implosions. Areal density asymmetries lead to a reduction in confinement, and so accurate measurement of such asymmetries is key in understanding current perturbation sources and improving experimental performance. 
	
	We developed a model for the singly scattered neutron spectrum which can fully account for the effects of an asymmetric areal density distribution. This required a more general description of the scattering geometry which defined the measurable neutron averaged areal density, $\langle \rho L \rangle$. The model outlined in this work will allow fits to spectroscopic data to infer the amplitude and mode number of the areal density asymmetries. The model was tested on neutron transport results for a simple mode 1 test case. Fitting the spectra from multiple lines of sight reproduced the theoretical results, even in the presence of non-uniform temperature and anisotropic birth spectra. This analysis provides the most direct measurement of areal density with little sensitivity to other factors. However, going to higher areal densities presents challenges due to the increasing level of multiple scattering.
	
	Due to the azimuthal symmetry of scattering, a single spectroscopic line of sight is insensitive to variation in $\langle \rho L \rangle$ along the azimuthal direction. Therefore, combining measurements from a sufficient number of detectors is required to obtain a complete picture of the $\langle \rho L \rangle$ distribution across the whole sphere. The error propagation for combining the measurements from multiple lines of sight was explored. This led to a methodology for calculating the optimal detector arrangement which minimised the error in the inferred physical quantities. This error analysis is also relevant to the measurement of primary DT and DD spectra which involve similar line of sight dependencies \cite{Hatarik2018,Mannion2020}.

	\section*{References}
	\bibliography{../MuCFRefs}
	
\end{document}